\documentclass[twocolumn]{aastex631}

\usepackage[T1]{fontenc}
\usepackage{CJK}

\DeclareRobustCommand{\VAN}[3]{#2}
\let\VANthebibliography\thebibliography
\def\thebibliography{\DeclareRobustCommand{\VAN}[3]{##3}\VANthebibliography}

\def\xHI{x_{\rm HI}}
\def\xHIv{\left<x_{\rm HI}\right>_{\rm v}}
\def\NHI{N_{\rm HI}}
\def\gbg{\Gamma_{\rm bkg}}

\usepackage{graphicx}	
\usepackage{amsmath}	

\begin{document}
\begin{CJK*}{UTF8}{gkai}

\title[CROC Mean Free Path]{Cosmic Reionization on Computers: The Evolution of Ionizing Background and Mean Free Path
}

\correspondingauthor{Jiawen Fan}
\author[0009-0002-9477-886X]{Jiawen Fan（樊稼问）}
\email{johnfan@umich.edu}
\affiliation{Department of Physics \\ University of Michigan, \\ Ann Arbor, MI 48109, USA}

\correspondingauthor{Huangqing Chen}
\author[0000-0002-3211-9642]{Huanqing Chen}
\email{hqchen@cita.utoronto.ca}
\affiliation{Canadian Institute for Theoretical Astrophysics\\
University of Toronto\\
Toronto, ON M5R 2M8, Canada
}

\author[0000-0001-8868-0810]{Camille Avestruz}
\affiliation{Department of Physics; Leinweber Center for Theoretical Physics\\
University of Michigan \\ Ann Arbor, MI 48109, USA
}

\author{Affan Khadir}
\affiliation{Department of Physics \\ University of Toronto \\ Toronto, ON M5R 2M8, Canada
}

\date{Accepted XXX. Received YYY; in original form ZZZ}

\begin{abstract}
Observations of the end stages of reionization indicate that at $z\approx 5-6$, the ionizing background is not uniform and the mean free path (MFP) changes drastically. As MFP is closely related to the distribution of Lyman Limit Systems and Damped Lyman-alpha Systems (LLSs and DLAs, or ionizing photon ``sinks''), it is  important to understand them. In this study, we utilize the CROC simulations, which have both sufficient spatial resolution to resolve galaxy formation and LLSs alongside a fully coupled radiative transfer to simulate the reionization processes.  In our analysis, we connect the evolution of the ionizing background and the MFP.  
We analyze two CROC boxes with distinct reionization histories and find that the distribution of ionizing background in both simulations display significant skewness 
that deviate from log-normal.
Further, the ionizing background in late reionization box still displays significant fluctuations ($\sim 40\%$) at $z\approx5$.
We also measure the MFP along sightlines that start 0.15 pMpc away from the center of potential quasar hosting halos.  The evolution of the MFP measured from these sightlines exhibits a break that coincides with when all the neutral islands disappear in the reionization history of each box (the `ankle' of the reionization history of the box).
In the absence of LLSs, the MFP will be biased high by $\approx 20\%$ at $z\approx 5$. We also compare the MFP measured in random sightlines.  We find that at $z\approx 5$ the MFP measured in sightlines that start from massive halos are systematically smaller by $\approx 10\%$ compared with the MFP measured in random sightlines.  We attribute this difference to the concentration of dense structures within 1 pMpc from massive halos. Our findings highlight the importance of high fidelity models in the interpretation of observational measurements.
\end{abstract}




\section{Introduction}

In the first billion years after the Big Bang, the intergalactic medium (IGM) in our universe experienced a major change called reionization. The ionizing photons emitted by the first generation of galaxies and quasars ionized neutral atoms, turning the IGM from a mostly neutral state to ionized state. Our understanding of this Epoch of Reionization comprises a fundamental piece in our understanding of the evolution of our Universe.


One pivotal question in reionization is the \textit{timeline of reionzation}. Although measurements of the cosmic microwave background have provided a strong constrain on the midpoint of reionization \citep[e.g.,][]{planck2020,spt2012,act2011}, it is still very uncertain if reionization occurred rapidly or if it prolonged over a long period of time \citep[e.g.,][]{paoletti2020,heinrich2021,naidu2020}. 
Recent findings from quasar spectra suggest that reionization ends later than we previously thought \citep[e.g.,][]{choudhury2015, weinberger2019, kulkarni2019}. In particular, multiple statistics from quasar absorption spectra indicate that the reionization process might have prolonged after $z=6$. These measurements include the distribution function of Lyman $\alpha$ optical depth \citep[e.g.,][]{bosman2022}, dark gap statistics \citep[e.g.,][]{becker2015,zhu2021,zhu2022,gnedin2022}, and measurements of the mean free path (MFP) \citep{Becker2021,bosman2021, Zhu2023,Satyavolu2023,roth2023, davies2023}.
These studies offer strong evidence that well below $z=6$, the ionizing background still retains large fluctuation or there may still be neutral IGM patches.

With the increasing evidence suggesting tension between observation and a simple uniform ionizing background model at $z\approx 6$, it is crucial to improve the models of the ending stages of reionization. It is widely acknowledged that ``sinks'' play an important role during reionization \citep{Gendin2006,Alvarez2012,Sobacchi2014,Park2016,Nasir2021}. Sinks largely correspond to Lyman Lymit Systems (LLSs) or Damped Lyman$\alpha$ Absorbers (DLAs) that stop the propagation of ionizing photons. These systems usually have scales under $\sim10$ kpc, while the majority of the simulations used to interpret observational data are semi-numerical with resolution above this scale \citep[e.g.,][]{davies2016}. 

At the same time, high-resolution radiative transfer cosmological hydrodynamic simulations are developed to understand the detailed physics during reionization, including CROC \citep{Gnedin14}, a spatially adaptive radiation-hydrodynamical simulation by \citet{pawlik2015}, THESAN \citep{Kannan2022}, SPHINX \citep{Rosdahl2018}, and CoDa \citep{Ocvirk2016}. With the resolution of $\sim 100$ pc and radiative transfer fully-coupled to gas dynamics, these simulations allow us to understand how ionizing radiation propagates in the universe ubiquitous with small dense structures better.

In this paper, we analyze two CROC boxes to understand how the ionizing background and MFP evolve. We focus on evolution at the tail end of reionization. This paper is organized as follows. We describe the CROC simulations, our methods to measure MFP, and the detection of LLSs in Section~\ref{sec:methods}, present ionizing background and mean free path measurements in Section~\ref{sec:results}, and summarize our conclusions in Section~\ref{sec:conclusions}.

\section{Methodology}\label{sec:methods}
\subsection{Simulations }\label{sec:meth:CROC Simulation}
We use Cosmic Reionization on Computers (CROC) Simulations to model the tail end of the epoch of reionization.
CROC simulations are run with Adaptive Refinement Tree code \citep{Kravtsov99,Kravtsov02,Rudd08} and include gravity, gas dynamics, fully-coupled radiative transfer, atomic cooling and heating, star formation, and stellar feedback. These implementations account for the physics necessary to accurately model of self-consistent cosmic reionization. The high resolution ($\sim 100$ pc) reasonably resolves dense structures such as Lyman Limit Systems (LLSs) and Damped Ly$\alpha$ absorbers (DLAs).  Full details of the simulation can be found in \citet{Gnedin14}. 

\begin{figure}
    \centering
\includegraphics[width=0.48\textwidth]{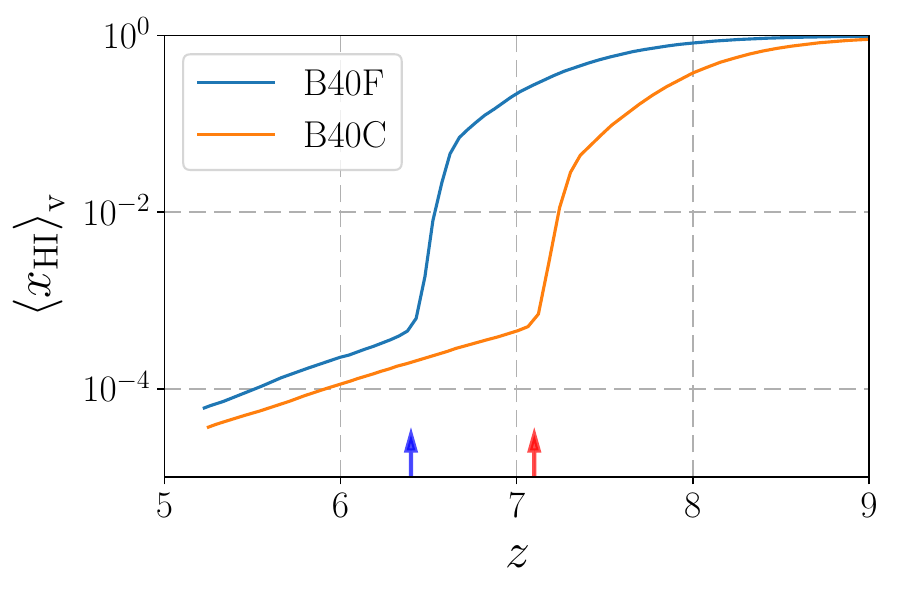}
\caption{The evolution of volume-weighted hydrogen neutral fraction in CROC B40F and B40C runs. Respectively, these are the latest and earliest reionized boxes in medium-sized CROC simulations.  Each respectively reaches the `end-stage' of reionization with all the neutral patches reionized by $z\sim6.4$ and $z\sim7.1$, the `ankle' point where we annotate with an arrow.}\label{fig:reihist}
\end{figure}

 In this paper, we analyze two simulation runs: B40F and B40C. Both have a box length $\rm L_{box}$ = 40 $h^{-1}$cMpc on each side. The two runs are performed with the exact same code and only differ in initial conditions. The DC mode \citep{gnedin2011,sirko2005} of B40F is $\Delta_{\rm DC}=-0.34$, corresponding to slightly underdense environments. The DC mode of B40C is $\Delta_{\rm DC}=0.50$, corresponding to slight overdense environments. This difference leads to a slightly earlier reionization in B40C than B40F (Figure \ref{fig:reihist}). We focus on these two runs because they represent the earliest and latest reionized $\rm L_{box}$ = 40 $h^{-1}$cMpc runs in the current available CROC products. Note that $\xHIv$ in both boxes drop quickly from $0.1$ to $0.001$, while the evolution afterwards is slow. This breaking point corresponds to where neutral patches in the IGM disappears.  We refer to the breaking point `ankle', corresponds to $z\approx6.4$ in B40F and $z\approx7.1$ in B40C, and refer to the time period after the breaking point ``end-stage'' of reionization.

\subsection{Sightline Generation and LLS identification}\label{sec:meth:los}\label{sec:meth:LLS_detection}

\subsubsection{Sightlines}
In each snapshot, we draw 1000 randomly-directed sightlines centered on the $20$ most massive halos identified by the \texttt{ROCKSTAR} halo finder \citep{Behroozi2013}. 
At $z=5.2$, the mass range of the most massive $20$ halos is $(0.4 \sim 1.1) \times 10^{12} \rm M_{\odot}$ in B40F and $(0.7 \sim 1.7) \times 10^{12} \rm M_{\odot}$ in B40C.
Note, our choice to
center sightlines on massive halos is to mimic the density environments of `quasar-like' sightlines. Hereafter in this paper, we use ``quasar-like'' to refer their mass and environment properties only; we do not consider quasar radiation in this work. 

We use the \texttt{yt} v3 \citep{Turk2011} function ``make\_light\_ray'' to generate sightlines for this study. Utilizing periodic boundary conditions, we generate $50$ pMpc long sightlines. We treat subregions of the sightlines that start $>15$ pMpc away from the halo centers as equivalent to sightlines centered on random positions in the box.

\subsubsection{LLS Identification}
LLSs and DLAs play an important role in determining the mean free path (MFP) in the later stages of reionization.  These are structures with greater than unit optical depth at the Lyman Limit. We therefore identify and characterize LLSs and DLAs in each sightline using the following procedure. The historic definition of LLSs and DLAs can be somewhat arbitrary: LLSs are absorbers with column density $N_{\rm HI}>1.6 \times 10^{17} \rm cm^{-2}$ and less than about $2 \times 10^{20} \:\rm cm^{-2}$ and DLAs are systems with column density greater than $2 \times 10^{20} \:\rm cm^{-2}$\citep{Crighton19}.

In this work, we follow \citet{jfan2024} to identify LLSs and DLAs that prove to work well in the CROC simulations: we use a neutral fraction threshold $x_{\rm HI} = 0.001$ to identify any high neutral fraction regions. We then integrate over the region with $x_{\rm HI} > 0.001$ using the number densities of neutral hydrogen $n_{\mathrm{HI}}$ along the sight lines.

\begin{figure*}
\centering
    \includegraphics[width=0.95\textwidth]{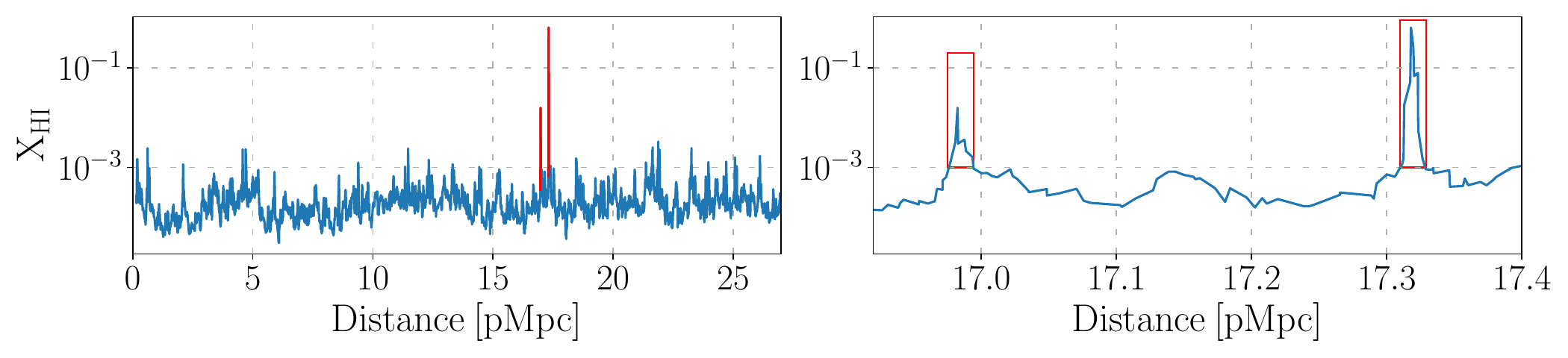}
    \caption{{Example LLSs identified in a sightline} from B40F at z $\sim$ 6. The left panel shows the neutral fraction of the sightline with the identified LLSs colored in red. The right panel zooms in to show the detected LLSs. These LLSs have column densities of $\log_{10}(\mathrm{N_{HI}[\mathrm{cm^{-2}}]}) = 17.3$ and $\log_{10}(\mathrm{N_{HI}[\mathrm{cm^{-2}}]}) = 19.5$.}
\label{fig:lls_visualization}
\end{figure*}

After calculating the column densities of the high neutral fraction regions, we select any structures with $N_{\rm HI}>1.6 \times 10^{17} \rm cm^{-2}$ to be LLSs/DLAs. Note, we use this criterion only for the late-stage of reionization, where there are no neutral IGM patches in the box (see Figure~\ref{fig:reihist}).

The left hand side of Figure~\ref{fig:lls_visualization} shows an example of LLSs detected along a sightline highlighted in red, and the right hand side of the panel shows zoomed-in part of the sight line. The LLSs detected are boxed in the red square. The two detected systems have column densities of $\log_{10}(\mathrm{N_{HI}[\mathrm{cm^{-2}}]}) = 17.3$ and $\log_{10}(\mathrm{N_{HI}[\mathrm{cm^{-2}}]}) = 19.5$.

\subsection{Removing LLSs/DLAs}\label{sec:meth:removing_LLS}


 After identifying and characterizing the LLSs in all sightlines, we explore how their presence affects MFP measurements. To this end, we compare the MFP in sightlines with LLSs/DLAs excluded and retained with results in Section~\ref{sec:results:aroundqso}. Note, LLSs can not be `physically' separated in any observational measurement of MFP given the definition Eq. ~\ref{eq:simple_mfp} provided above. However, the high fidelity simulation enables the selection and exclusion of such structures in our measurement of MFP. We exclude simulation cells in which we have detected LLSs, effectively extracting an LLS-free sightline from which we can measure the MFP `without LLSs/DLAs'.

\section{Results}\label{sec:results}
\subsection{Ionizing Background Evolution}\label{sec:gamma_bkg_calculation}

The background ionization rate of neutral hydrogen, $\gbg$, is closely related to the mean free path of ionizing photons, $\lambda_{\mathrm{mfp}}$. However, non-uniformly distributed ionizing sources and photon sinks cause fluctuations in the measurements of $\gbg$. With structure formation and the evolving neutral fraction during the epoch of reionization, we expect the $\gbg$ fluctuations to vary and potentially exhibit a complex structure.  In this subsection, we quantify the $\gbg$ evolution and distribution in CROC.

\begin{figure}
    \centering
\includegraphics[width=0.48\textwidth]{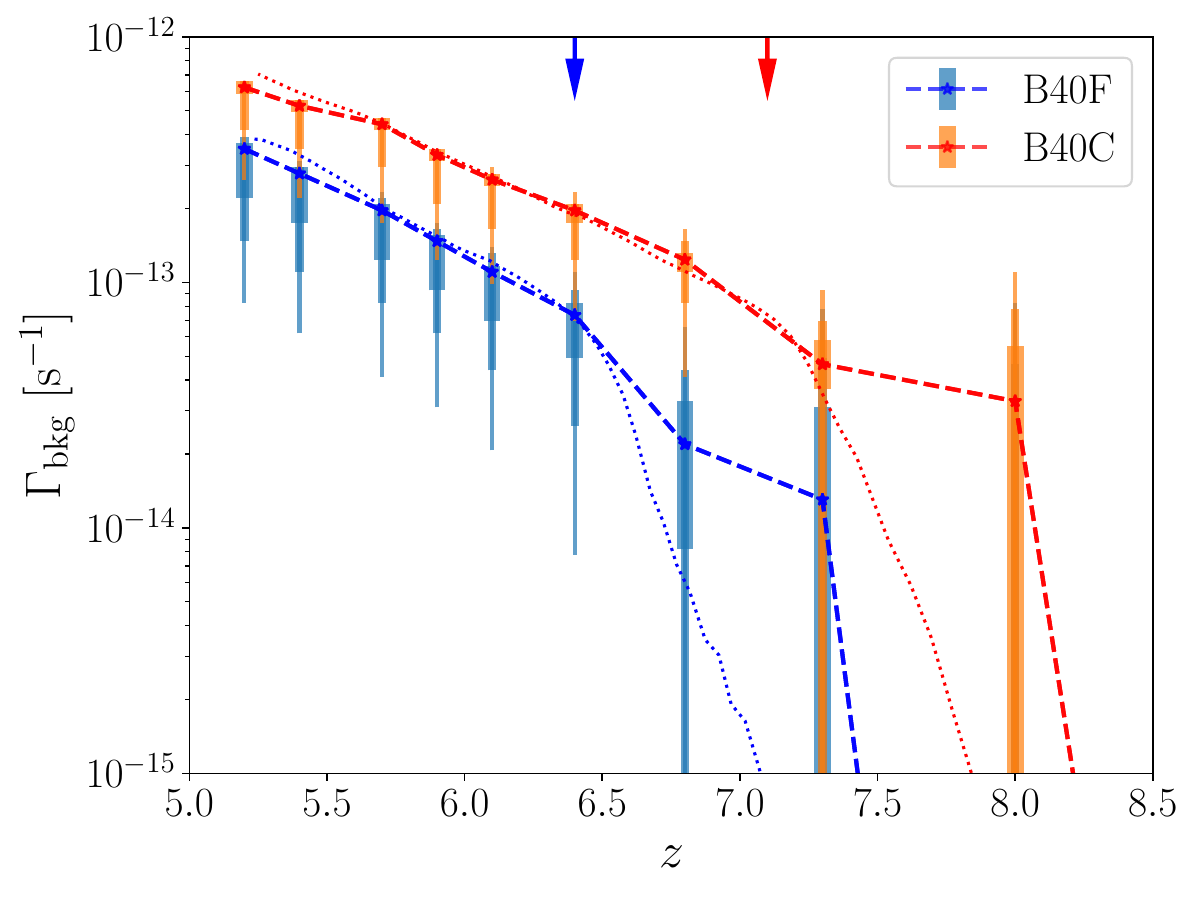}
    \caption{{Evolution of the Ionizing Background Radiation} We illustrate the estimated background ionization rate of neutral hydrogen as a function of redshift in B40F and B40C. The error bars show the $68\%$, $95\%$ and $99.7\%$ spread in the values estimated from uniformly sampling 1000 lines of sight. The later reionization box has a tighter distribution of ionizing background radiation at later times (orange errorbars at low-z). Dotted lines show the spatially averaged $\gbg$ calculated exactly on-the-fly.
}    
    \label{fig:gamma_evo}
\end{figure}

The ionizing background $\gbg$ is not a saved field in the CROC simulation products and is complicated to recalculate. We instead use the saved quantities (e.g., neutral fraction, temperature) as a proxy to calculate $\gbg$, assuming ionization equilibrium

:

\begin{equation}\label{eq: bkg_gamma}
\frac{d x_{\mathrm{HI}}}{d t}=-\left(\gbg + n_{e} \Gamma_{e}\right)x_{\mathrm{HI}} + \alpha^{\mathrm{HI}}(T) n_{e}{x_{\rm HII}}=0.
\end{equation}
Here, $\Gamma_{e}$ is the collisional ionization rate and $\alpha^{\mathrm{HI}}$ is the recombination rate of $\mathrm{HI}$, both of which depends on gas temperature. Using Eq.~\ref{eq: bkg_gamma}, we calculate $\Gamma_{\mathrm{bkg}}$ in each cell that our sightlines cross. Note that strictly speaking, the IGM is not in a perfect ionization equilibrium, but the process is slow enough that this formula should give a good approximation. For comparison, in Figure~\ref{fig:gamma_evo}, the dotted lines show the spatially averaged $\gbg$ calculated exactly on-the-fly. Note, the spatially averaged $\gbg$ reasonably follows the evolution of median estimated $\gbg$ from sightline quantities after the disappearances of neutral patches (to the left of the arrow tickmarks for each respective box). 

Figure~\ref{fig:gamma_evo} shows the evolution of the estimated ionizing background radiation measured from properties extracted from 1000 lines of sight at each redshift snapshot.  Error bars of increasing width correspond to the $68\%$, $95\%$, $99.7\%$ spread in the measurement. We connect the median values with the dashed line of the same color. Overall, the background ionization rate increases with time (decreasing redshift). The blue data points correspond to the median measurement from lines of sight extracted from a box with a later reionization time, and the orange data points correspond to the median measurement from a box with an earlier reionization time.   At the redshifts when IGM neutral patches disappear, corresponding to $z\approx6.4$ for B40F and $z\approx 7.1$  for B40C, the median $\Gamma_{\mathrm{bkg}}\approx 10^{-13} \rm s^{-1}$. 
Note, the orange data points have a relatively tighter set of errorbars at lower redsfhift. In the next paragraph we look into details of the histogram of $\gbg$.

\begin{figure}
    \centering
\includegraphics[width=0.48\textwidth]{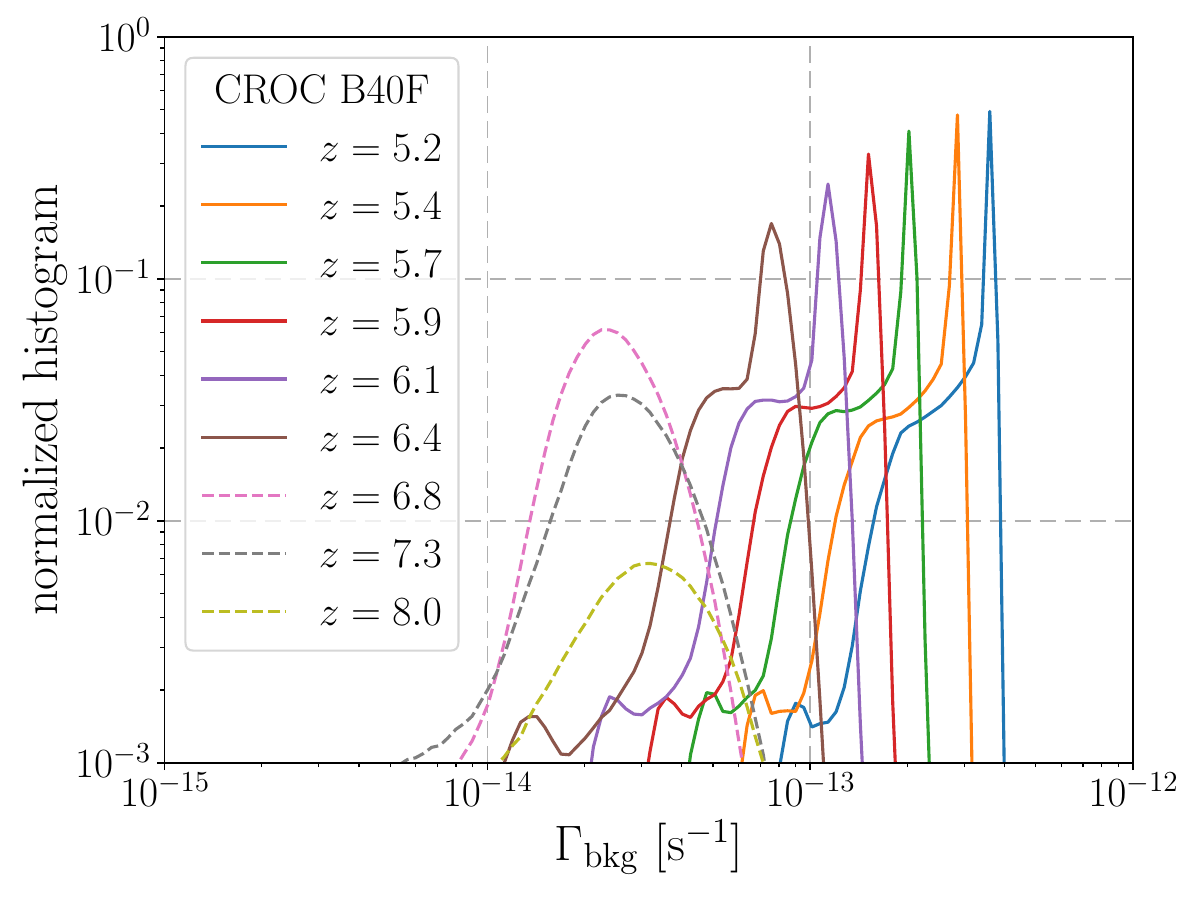}
    \caption{Distribution of the ionizing background $\gbg$ of B40F (underdense, late reionization box). The distribution is calculated by uniformly sampling the 1000 sightlines at each redshift.
    Here, we see a complex shape to the histogram, particularly at the low $\gbg$ tail.  Dashed lines correspond to the distribution of the ionizing radiation prior to the disappearance of neutral patches, or the `ankle' in the reionization history of the box.}
    \label{fig:gamma_scatter_box_F}
\end{figure}

\begin{figure}
    \centering
\includegraphics[width=0.48\textwidth]{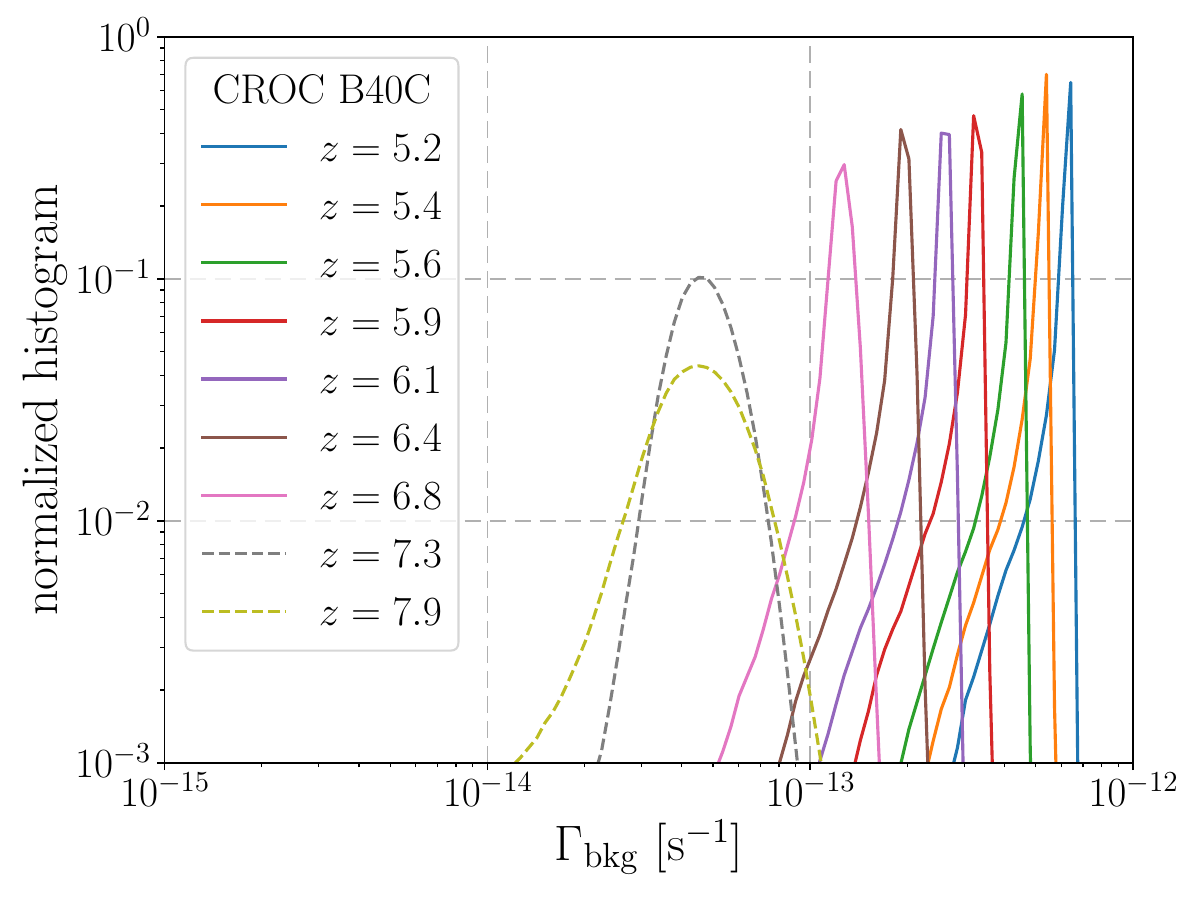}
    \caption{Same as Figure \ref{fig:gamma_scatter_box_F} but for B40C, the overdense, early ionization box. Here, we see a smoother and narrower distribution compared with the late reionization counterpart in Figure~\ref{fig:gamma_scatter_box_F}.   Dashed lines correspond to the distribution of the ionizing background radiation prior to the disappearance of neutral patches, or the `ankle' in the reionization history of the box.} 

    \label{fig:gamma_scatter_box_C}
\end{figure}

In Figure~\ref{fig:gamma_scatter_box_F} and Figure~\ref{fig:gamma_scatter_box_C}, we show the entire normalized $\gbg$  histogram evolution.  We show histograms for the $\gbg$ before the `ankle' of the reionization with dashed lines, and histograms after the `ankle' with solid lines. Before the `ankle', many regions have zero $\gbg$, which does not contribute to the histogram. This is also the reason why the dashed lines enclose a smaller areacannot be shown in under the curve.
The shapes of $\gbg$ distributions before the `ankle' of the reionization are largely consistent, following a near log-normal shaped distribution. 
For the $\gbg$ distributions after the `ankle', their peaks skew significantly towards larger $\gbg$. The evolution of the $\gbg$ histograms between the two boxes display similarities as well as distinct features. In both boxes, the shape of the $\gbg$ distribution stabilizes once the neutral patches disappear, with the entire distribution moving to the larger $\gbg$ values with time. However, the shape of the $\gbg$ after the `ankle' of reionization significantly differs between the two boxes. In the underdense and later reionizating box (B40F), the shape of the distribution exhibits more features, including a distinctive `bump' at the low $\gbg$ tail that persists to $z\approx 5$. On the other hand, the overdense and earlier reionizing box (B40C) exhibits a much less prominent `bump'. In the Appendix, we check another box B40E with reionization history in between B40F and B40C.  The `bump' feature also appears the box with intermediate reionization history compared with the other two. We attribute the tightness of the low redshift orange errorbars in Figure~\ref{fig:gamma_evo} to the fact that an early reionization box (with relatively overdense initial conditions) has more ionizing 
sources.  The excess of ionizing sources leads to higher $\gbg$ peaks in the distribution and suppresses smaller scale fluctuations in the ionizing background. In the Appendix we also show the $\gbg$ maps to support this explanation.

\subsection{Mean Free Path Measurements: Two Definitions}\label{sec:meth:Mean Free Path Calculation}

The mean free path of ionizing photon is defined as the average length one photon can travel before absorbed by neutral gas:
\begin{equation}\label{eq:simple_mfp}
   \left<\sigma_{\rm LyC} \lambda_{\mathrm{mfp}} n_{\rm HI} \right>= 1
\end{equation} 
However, we could define this quantity in more than one way that depends on the averaging procedure.  We highlight two distinct approaches.  The first is a simple definition where we ignore the frequency dependency, denoted as $\lambda_{\rm mfp, sim}$.  The second is an observationally-motivated approach, where we define this in the absorption spectra, denoted as $\lambda_{\rm mfp, obs}$.

{In the simple definition}, 
the frequency dependence of $\sigma_{\rm LyC}$ is ignored and the integrated $\left< \sigma_{\rm LyC}\right> =6.3\times 10^{-18} \rm\  cm^2$ is often used.
We measure the MFP defined this way by calculating $\lambda_{\mathrm{mfp}}$ of each sightline where 
\begin{equation}\label{eqn:def}
    \frac{1}{\left< \sigma_{\rm LyC} \right>} =  \int_0^{d}n_{\rm HI} dr
\end{equation}
and $\lambda_{\mathrm{mfp}}=\left< d \right>$ for all the 1000 sightlines. In our comparison, we denote the MFP defined this way as $\lambda_{\rm mfp,sim}$. Note that in practice, we cut out the inner $0.15$ pMpc before the start of integration. This is to avoid the dense gas inside the halo. Observationally, the MFP is measured from quasar sightlines, where the dense neutral gas inside the host halo has been cleared out by the quasar. Excluding the first $0.15$ pMpc mimics such effect. Similarly, when we making LyC absorption spectra described in the following paragraph, we also exclude the first $0.15$~pMpc. We chose to exclude this specific radius of $0.15$ pMpc because it is much larger than the largest halo in our simulation, which has a virial radius of $0.06$ pMpc. This ensures that all dense gas in the host galaxy does not affect our measurements.  We have also tested excluding a larger radius of $0.2$~pMpc from the halo center and found that the result does not significantly change.

{In an observavtionally-motivated approach,} 
the MFP is defined as the distance where the transmitted flux blueward of the Lyman continuum drops to $e^{-1}$ \citep{prochaska2009,Becker2021}. In our comparison, we denote the MFP defined this way as $\lambda_{\rm mfp,obs}$. First, we assess whether there are any significant differences between these two approaches to calculate the MFP. Using the simulation, we create mock spectra by convolving the LyC transmission after each gas cell along the sightlines. The opacity depth blueward of the Lyman continuum is defined as,
\begin{equation}\label{eq:trans_sim}
    \tau (\nu) =\int_0^{\infty} n_{\rm HI} \sigma_{\rm LyC}(\nu' -\nu) dr,
\end{equation}

where $r$ is the distance from the gas cell to the quasar, $\sigma_{\rm LyC}$ is the photoionization cross-section \citep{bolton2007}:
\begin{equation}\label{eq:xsec}
    \sigma_{\rm LyC}(\nu)= 6.3 \times 10^{-18}[1.34(\nu/\nu_{\rm 912})^{-2.99}-0.34(\nu/\nu_{\rm 912})^{-3.99}] \rm cm^{2}.
\end{equation}

Depending on the frequency corresponding to the gas cell along the sightlines, we have $\nu'$ defined as:
\begin{equation}
    \nu'=\nu_{\rm 912} (1+\frac{Hr-v_{\rm pec}}{c}).
\end{equation}
\begin{figure*}
    \centering
    \includegraphics[width=0.3\textwidth]{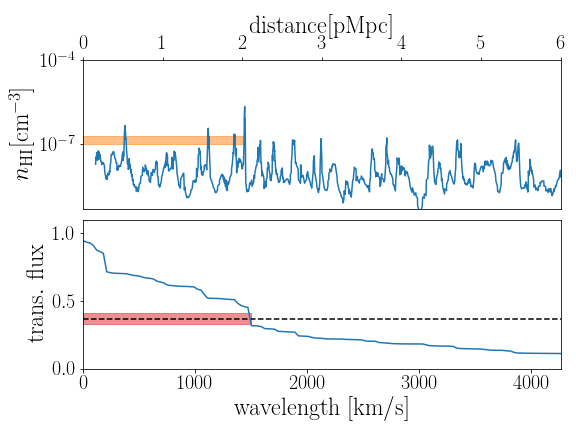}
    \includegraphics[width=0.3\textwidth]{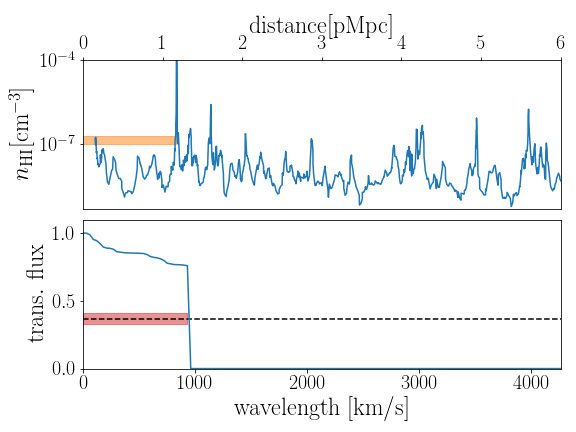}
        \includegraphics[width=0.3\textwidth]{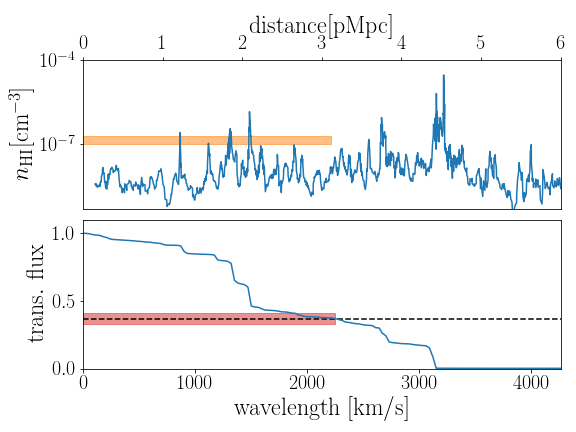}
    \caption{{Example sightlines comparing the two definitions of the MFP} for sightlines in B40F at $z=6.1$. Upper panels show the neutral hydrogen number density and the lower panels show the transmitted LyC flux. The horizontal line marks $e^{-1}$, the threshold used to define $\lambda_{\rm mfp, obs}$. The orange bar in the upper panel indicates the length of MFP defined using integrated cross section (``simple approach''), while the red bar in the low panels indicates the length of MFP defined using 1/e threshold in transmitted flux (``observer's approach''). They have similar length. In the majority of the cases, the MFP stop at structures of column densities $\log10(N_{\rm HI \rm [cm^{-2}]})=15.5-17$, while some sightlines hit LLSs where the MFP ends. Very occasionally, a sightline does not encounter any dense structures and has longer than average MFP (right panel).}
    \label{fig:lls_mfp_interaction}
\end{figure*}

\begin{figure}
    \centering
    \includegraphics[width=0.45\textwidth]{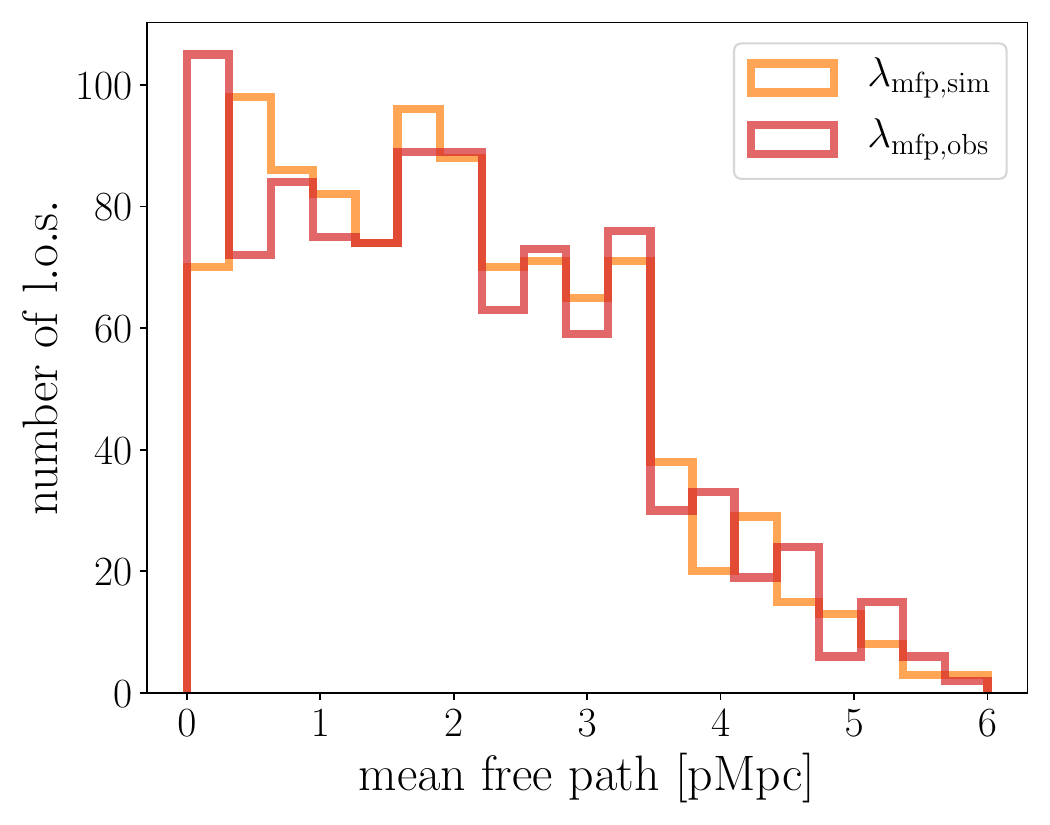}
    \caption{{Comparison of MFP distribution between different definitions.} Histograms of mean free path using two definitions stated in Sec. \ref{sec:meth:Mean Free Path Calculation}. We draw 1000 `quasar-like' sightlines from the B40F $z=6.1$ snapshot.  The distributions are similar.}
    \label{fig:compare_two_def}
\end{figure}


In Figure \ref{fig:lls_mfp_interaction}, we show some examples of the sightlines drawn from B40F at $z=6.1$. Upper panels show the neutral hydrogen number density and the lower panels show the transmitted LyC flux. The orange strips in the upper panels indicate the length of $\lambda_{\rm mfp, sim}$ and the red strips in the lower panels indicate the length corresponding to $\lambda_{\rm mfp, obs}$. {We find that the MFP of most sightlines stop at a ``sub-LLS'' ($\NHI<1.6\times10^{17} \rm cm^{-2}$), similar to the case in the left panel in Figure \ref{fig:lls_mfp_interaction}.
The MFP of most other sightlines stop at an LLS/DLA, similar to that in the middle panel. The case where the MFP stops at an extended voids (right panel) is relatively rare.} From examining individual examples like those in Figure \ref{fig:compare_two_def}, the mean free path defined in the two different ways are almost identical. In fact, the difference between the two definitions is less than $5\%$ for the vast majorities of the sightlines, confirming that the peculiar velocity and frequency dependency of $\sigma_{\rm LyC}$ is not important. In Figure~\ref{fig:compare_two_def}, we plot the histogram of the MFP with the two definitions. The histogram is based on 1000 sightlines drawn from the simulation snapshot B40F at $z=6.1$. The distribution of both MFP measurements are indistinguishable with a $p-$value$=0.11$ using the Kologorov-Smirnov test. 

Because these two definitions result in almost identical MFP values, we hereafter use the first definition defined in Equation~\ref{eqn:def} for consistency, where $\lambda_{\mathrm{mfp}} = \left<d\right>$ for the rest of the paper.

 \begin{figure}
     \centering
     \includegraphics[width=0.45\textwidth]{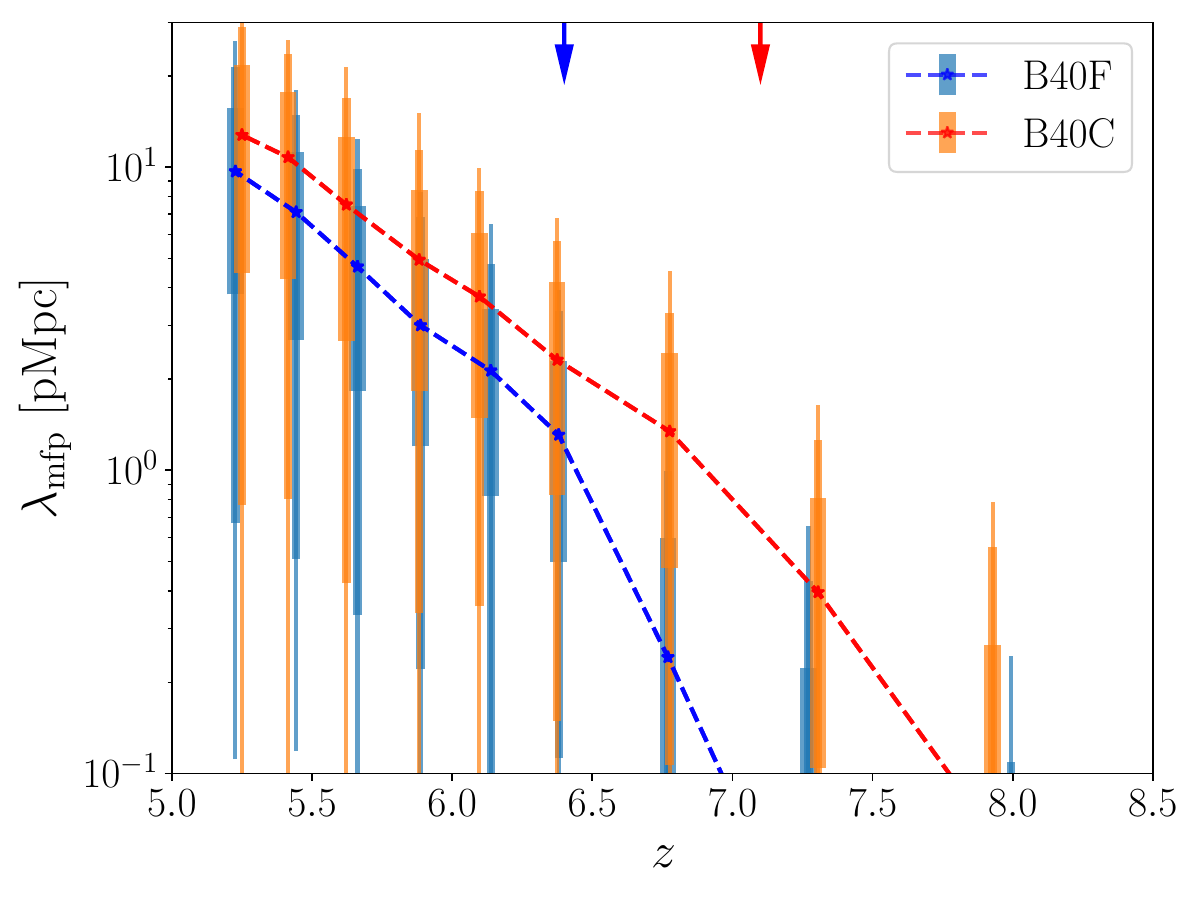}
     \caption{{Evolution of the mean free path.}  B40C has an earlier reionization epoch than B40F, and exhibits a systematically larger mean free path.  Data points correspond to the median value, connected by the dashed line.  The error bars show the $68\%$, $95\%$ and $99.7\%$ scatter.}
     \label{fig:mfp_evo}
 \end{figure}

\begin{figure}[bt!]
\centering
        \includegraphics[width=0.9\columnwidth]{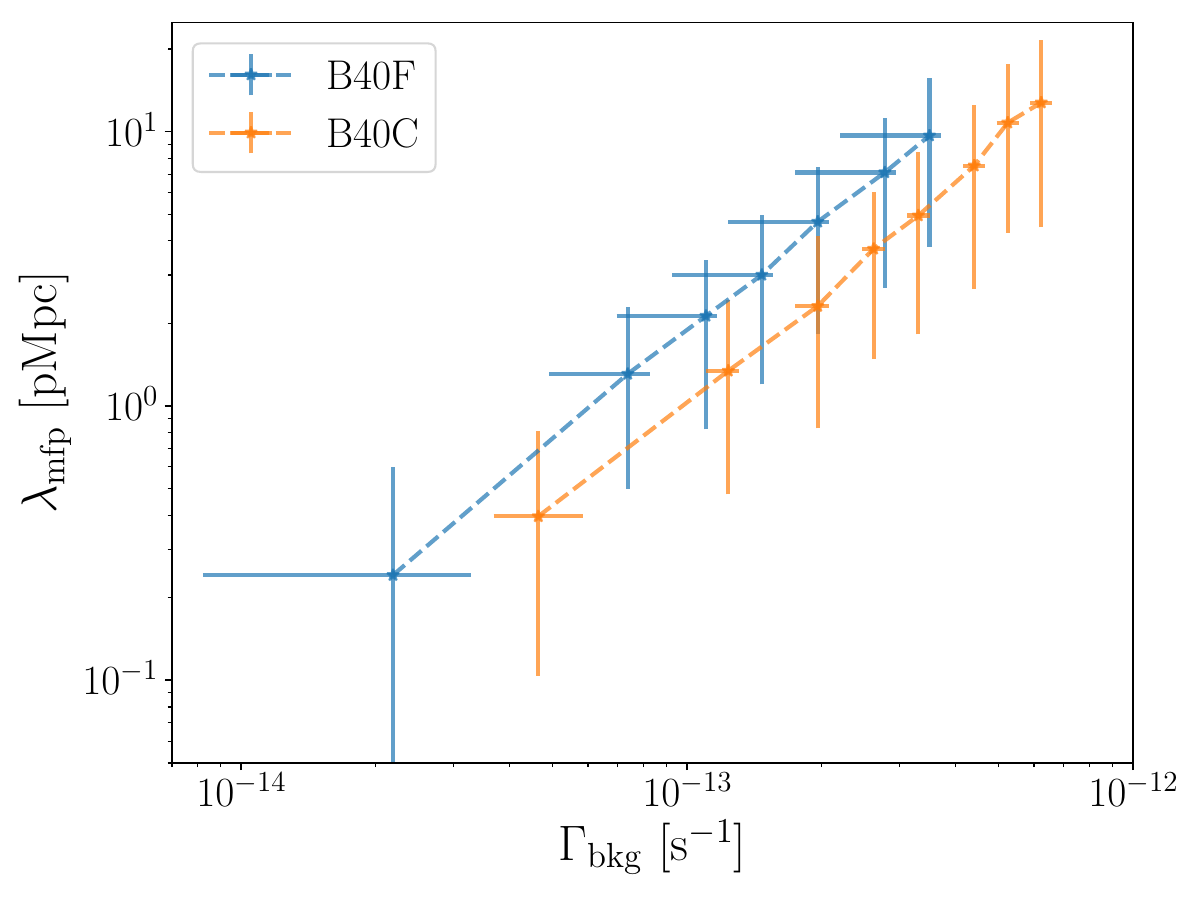}
\caption{Evolution of Mean Free Path over Ionizing Background Radiation for both box B40F and B40C. The redshift ranges from the snapshot just before the `ankle' point in each box (lower leftmost errorbar data point) and each snapshot to $z \approx 5.2$. For both boxes, the two quantities follow a power-law relation of the same slope but with offset.
}
\label{fig:mfp_vs_gamma}
\end{figure}

\begin{figure*}
\centering
\includegraphics[width=0.49\textwidth]{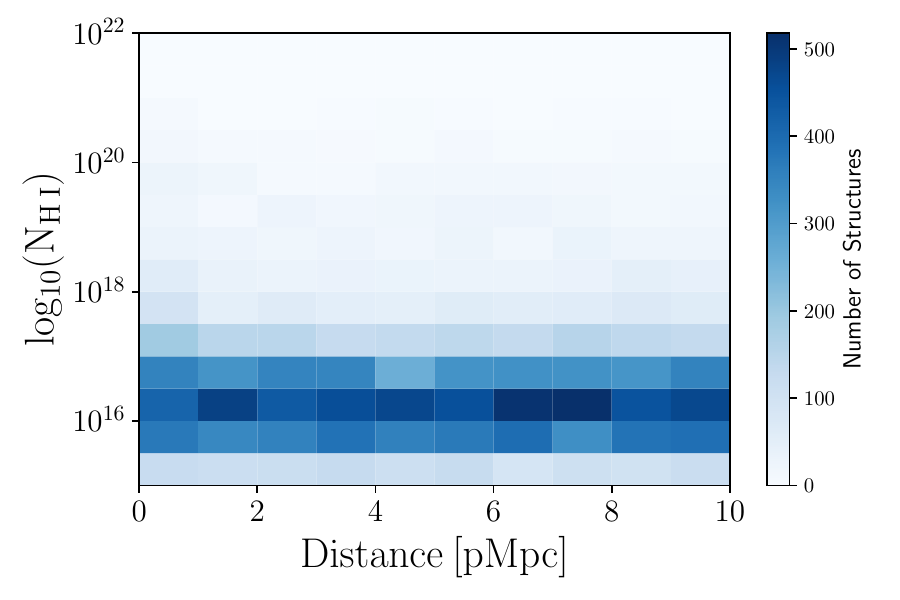}
\includegraphics[width=0.49\textwidth]{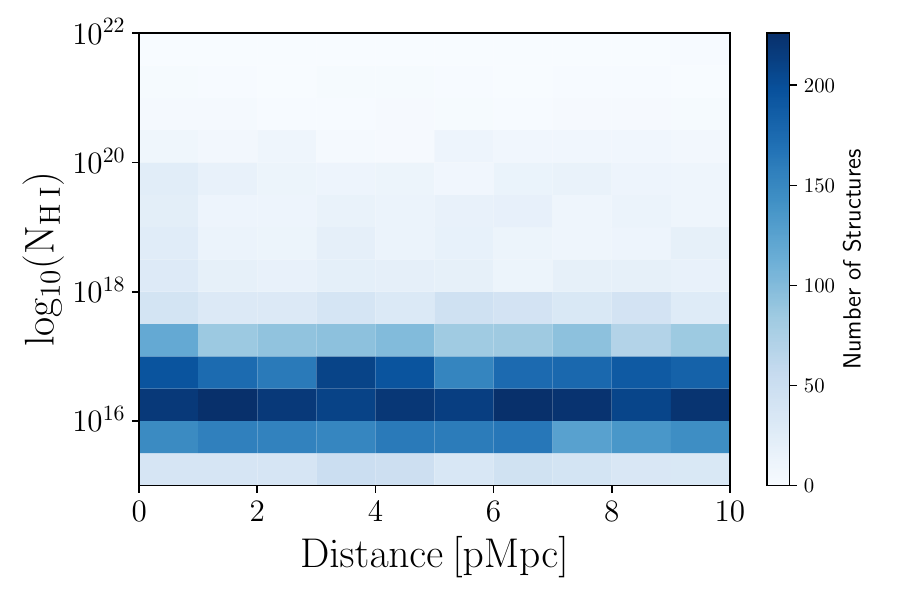}
\includegraphics[width=0.49\textwidth]{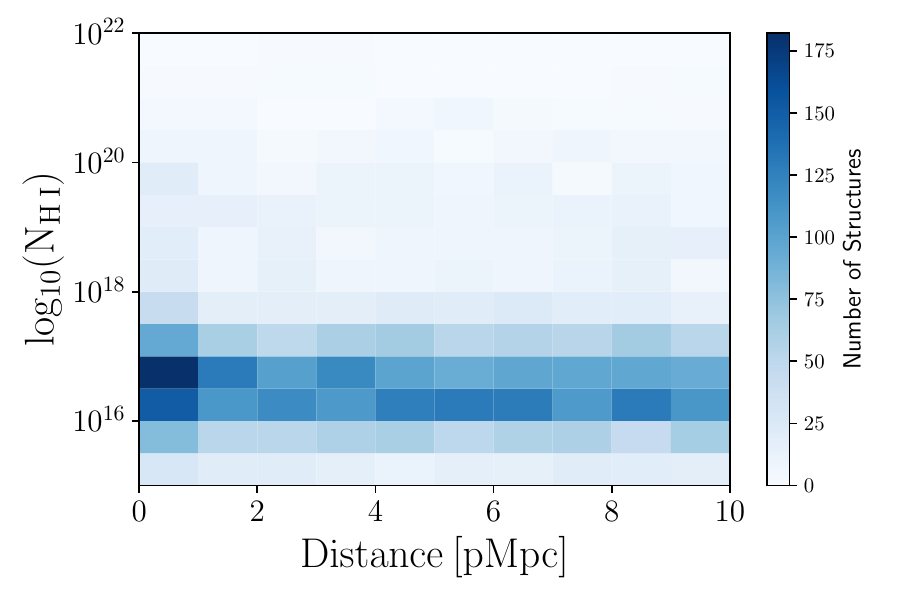}
  \includegraphics[width=0.49\textwidth]{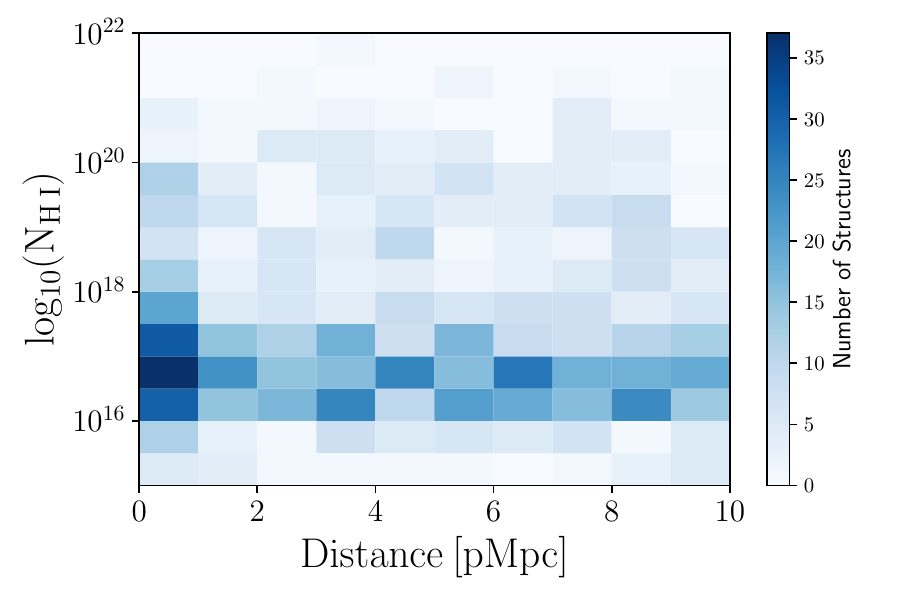}
\caption{From left to right and top to bottom, 2D number counts of structures in all 1000 sightlines at $z=6.4, 6.1, 5.9, 5.2$ in B40F using the $x_{\rm HI}>10^{-3} \rm cm^{-3}$ criteria described in Section~\ref{sec:meth:LLS_detection}. At $z>6$, there are more clumpy structures at $\log_{10}{\NHI} [{\rm cm^{-2}}] =15.5 - 17$ because the ionizing background is lower and ionization time is shorter. Afterwards, most structures at $>1$ pMpc disappear after sustaining increasing ionizing background for a longer time. However, a relative concentration of dense clumps within $>1$ pMpc from the center of massive halos appears.} 
\label{fig:column_density_vs_distance_evo}
\end{figure*}

\begin{figure*}
\centering
        \includegraphics[width=0.4\textwidth]{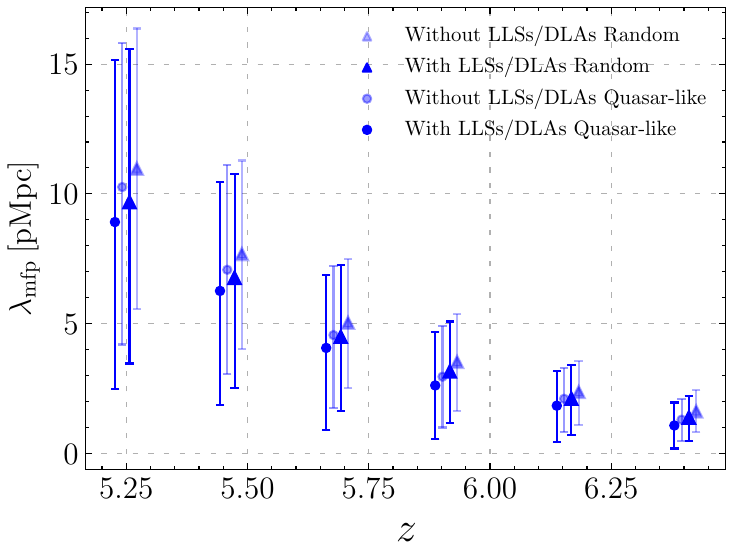}
        \includegraphics[width=0.4\textwidth]{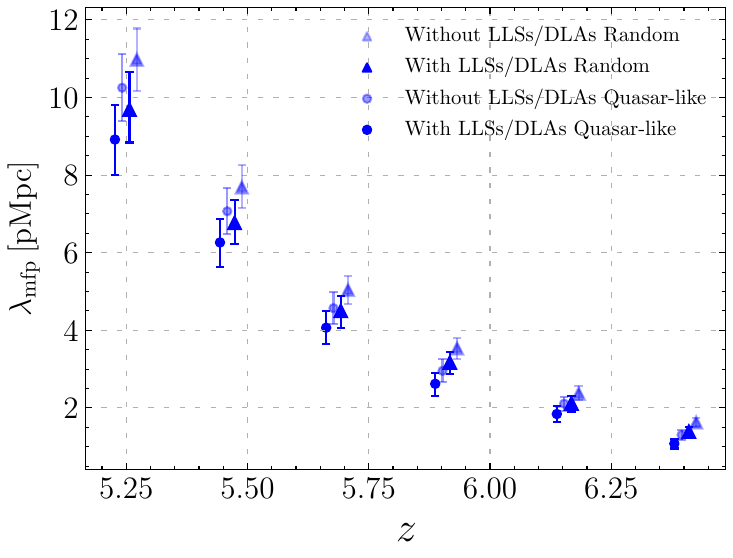}
\caption{MFP evolution in B40F. We vary choices in our MFP measurements using the same sightlines. The data points at the same redshift are offset for clarity.  The blue solid circle indicates the value of MFP measured from quasar-like sightline starting points. The blue solid triangle indicates the value of MFP measured from random starting locations in the box. The transparent circle and triangle indicate MFP measured after removing any structures above $N_{\rm HI}>1.6 \times 10^{17} \rm cm^{-2}$ using the method mentioned in Sec~\ref{sec:meth:removing_LLS} along sightlines in both quasar-like sightlines and random locations respectively. The error bars on the left panel show the $68\%$ scatter for individual sightlines. The error bar in the right panel shows the $68\%$ scatter of the measured MFP mean, obtained through bootstrap resampling of 40 individual sightlines from the random 1000 sightlines.}
\label{fig:mfp_different_environments}
\end{figure*}

\subsection{Evolution of Mean Free Path}
In Figure \ref{fig:mfp_evo},  we show the evolution of the mean free path in both the CROC boxes with an late and a early reionization history. The labels and color style are similar to Figure \ref{fig:gamma_evo}, where the blue denotes B40F (late) while the orange (early) denotes B40C. 
The dashed lines show the median evolution of the MFP. The thick, medium, and thin errorbars show the $68\%$, $95\%$ and $99.7\%$ scatter of the MFP. The MFP of sightlines from the late reionization box (blue curve) is smaller than that of that of the early reionization box (orange curve) at all times. 

At the $z=6.4$, the slope of the median MFP in B40F exhibits a noticeable break in slope when the median MFP evolution transitions from a steep increase at early times and shallower increase at later times.  The epoch of the break coincides with when the reionization history of B40F exhibits a drastic change at $\xHIv\approx 10^{-3}$ (Figure \ref{fig:reihist}), or the `ankle' of the B40F reionization history (denoted by the blue arrow tick mark at $z=6.4$). The break in the B40C median MFP curve from the early reionization box is as apparent. While the break could be present in the evolution of the B40C MFP, it may not be as prominent due to the limited sampling above $z=7$.

We also examine the relationship between the MFP and the ionizing background radiation.  In Figure~\ref{fig:mfp_vs_gamma}, we show how these two quantities scale with one another.  In general, as the redshift decreases and reionization approaches completion, the median $\gbg$ increases with the median MFP.  Both boxes exhibit a similar slope ($\approx 1.3$) in scaling. 
However, the late reionization box B40F shows a systematically larger MFP at fixed $\gbg$ compared with that of the early reionization box B40C. The difference in normalization can be explained by the larger fluctuations in B40F. Because these `quasar-like' sightlines start from massive halos, the B40F sightlines center on slightly higher $\gbg$ regions compared to those from B40C. We also noticed that the data point just before the `ankle' (the leftmost data point) in both boxes obey the same powerlaw scaling. 
The obeyance of the same scaling relation for this data point before the ``ankle'' is due to the selection effect of the sightline environments, which start from massive halos.  The massive halos typically reside in large ionized bubbles, mimicking the behavior of an already reionized box.

\subsection{Environment Effects on Mean Free Path}\label{environments_vs_mfp}

Because most LLSs exist near halos \citep{jfan2024} and halos cluster together, we expect measurements of the MFP that start from halos (`quasar-like' sightlines) to bias low compared with random sightlines.  The high-resolution CROC simulations provide a solid test ground to investigate how much the environment of quasar-like sightlines and LLSs contribute to such biases in measurements of the MFP.

To prevent dense gas inside the host halo from contributing to our MFP measurements, we exclude this region from our `quasar-like' sightlines. We cut 0.15 $\mathrm{pMpc}$ from the start of every sightline, which is a sufficient distance from the center of the largest dark matter halos. Note, the largest dark matter virial halo radius in our catalog is  $0.06 ~\mathrm{pMpc}$. \footnote{We have checked the dependence of the median MFP calculation on the starting point from the halo center.  The MFP significantly decreases if it starts within 0.05 $\mathrm{pMpc}$ within the halo center, motivating our choice of 0.15 $\mathrm{pMpc}$ to be far enough.}
 
\subsubsection{Regions Around Quasar}\label{sec:results:aroundqso}

In Figure ~\ref{fig:column_density_vs_distance_evo}, we show the number of dense structures in our `quasar-like' sightlines that cross the $\xHI>10^{-3}$ threshold in B40F, the late reionization box. 
The color bar of the 2-D histogram shows the number counts of LLSs and DLAs detected along our sightlines, binned by their column density and their distance from the start of our `quasar-like' sightline. We use the method mentioned in Sec.~\ref{sec:meth:LLS_detection} to calculate the column density of detected LLSs and DLAs. 

The top panels illustrate the distribution for two epochs at and just after the `ankle' in the box reionization history, $z=6.4$ (top left) and $z=6.1$ (top right), respectively.  The bottom panels illustrate the same distribution for two epochs far after the `ankle', $z=5.9$ (bottom left) and $z=5.2$ (bottom right).  At this point, even the last neutral patches have been reionized for $>100$ Myr.  At the `ankle', there are a significant number of structures with $\NHI=1\times10^{16} \sim 1\times10^{17} \rm cm ^{-2}$ at all distances along the sightline. As reionization proceeds, most of such structures quickly disappear.  At $z=5.2$ (lower right panel), there persists a concentration of 
dense structures within first $\sim1~\rm Mpc$ of the quasar host halo center. After the `ankle' of the box reionization history, the likelihood of a `quasar-like' sightline encountering a LLSs/DLAs ($\NHI>1.6\times 10^{17}\rm cm^{-2}$) within the first 1 pMpc is significantly higher than a sightline that starts from a random position.

\textit{Dependence of MFP on sightline starting point:} In Figure~\ref{fig:mfp_different_environments}, we quantify this environmental effect by comparing the MFP of `quasar-like' sightlines (circles) to those that start at random positions (triangles). To calculate the MFP from a random location, we measure the MFP from a random starting position on the sightline between $15-25$~pMpc away from the halo center instead of the $0.15$~pMpc starting distance for our `quasar-like' MFP measurement. We measure the MFP for quasars across 1000 random lines of sight at each redshift. The left panel of Figure~\ref{fig:mfp_different_environments} shows the mean of this measurement in the dark circles and we find that the MFP value measured from random locations at each redshift is systematically higher by $\approx 10\% \sim 20 \%$. While this bias is much smaller than the scatter of MFP across individual sightlines, the bias could be important when we analyze a large sample of sightlines ($\gtrsim 40$, see the right panel illustrating a bootstrap resampling of 40 sightlines).

\textit{Dependence of MFP on presence of LLSs/DLAs:} We additionally use the method described in section~\ref{sec:meth:removing_LLS} to remove dense structures along sightlines that correspond to LLSs/DLAs.  As apparent in the progression from $z=6.4$ to $5.2$ in Figure~\ref{fig:column_density_vs_distance_evo}, dense structures column density ($N_{\mathrm{HI}}> 10^{17}$) tend to disappear further from the start of the `quasar-like' sightlines due to the increasing ionizing background, leading to a clustering of dense neutral hydrogen clouds near dark matter halo centers. When removing LLSs/DLAs, the MFP noticeably increases for `quasar-like' sightlines. For example, at the lowest redshift of $z\approx5.2$ shown in Figure~\ref{fig:mfp_different_environments}, we see the mean MFP shift from $\lambda_\mathrm{mfp}\sim8.9$~pMpc (dark blue circle) to $\lambda_\mathrm{mfp}\sim10.5$~pMpc (light blue circle).  For the measurement in random sightlines, the relative increase in mean MFP due to dense structure removal is somewhat more modest, $\lambda_\mathrm{mfp}\sim9.5$~pMpc (dark blue triangle) to $\lambda_\mathrm{mfp}\sim10.8$~pMpc (light blue triangle). At redshift $z \approx 5.2$, the removal of dense structures along sightlines leads to an $\approx 20 \%$ increase in the mean measured MFP. 
In the right panel of Figure ~\ref{fig:mfp_different_environments}, we show the mean and $68\%$ scatter of bootstrap resampling mean of 40 randomly selected sample sightlines from the original 1000 sightlines at each redshift.
We observe that the scatter of mean MFP of this sample size is smaller than the difference between MFP measured starting from random locations without LLSs/DLAs (faint triangles) and MFP measured starting from `quasar-like' halos with LLSs/DLAs (dark blue circle). This shows that the combination of environmental factors and small-scale structures can become important in measuring MFP when the sample size reaches $\gtrsim 40$.

\subsection{Comparison With Observational MFP Constraints}
\begin{figure}
    \includegraphics[width=0.45\textwidth]{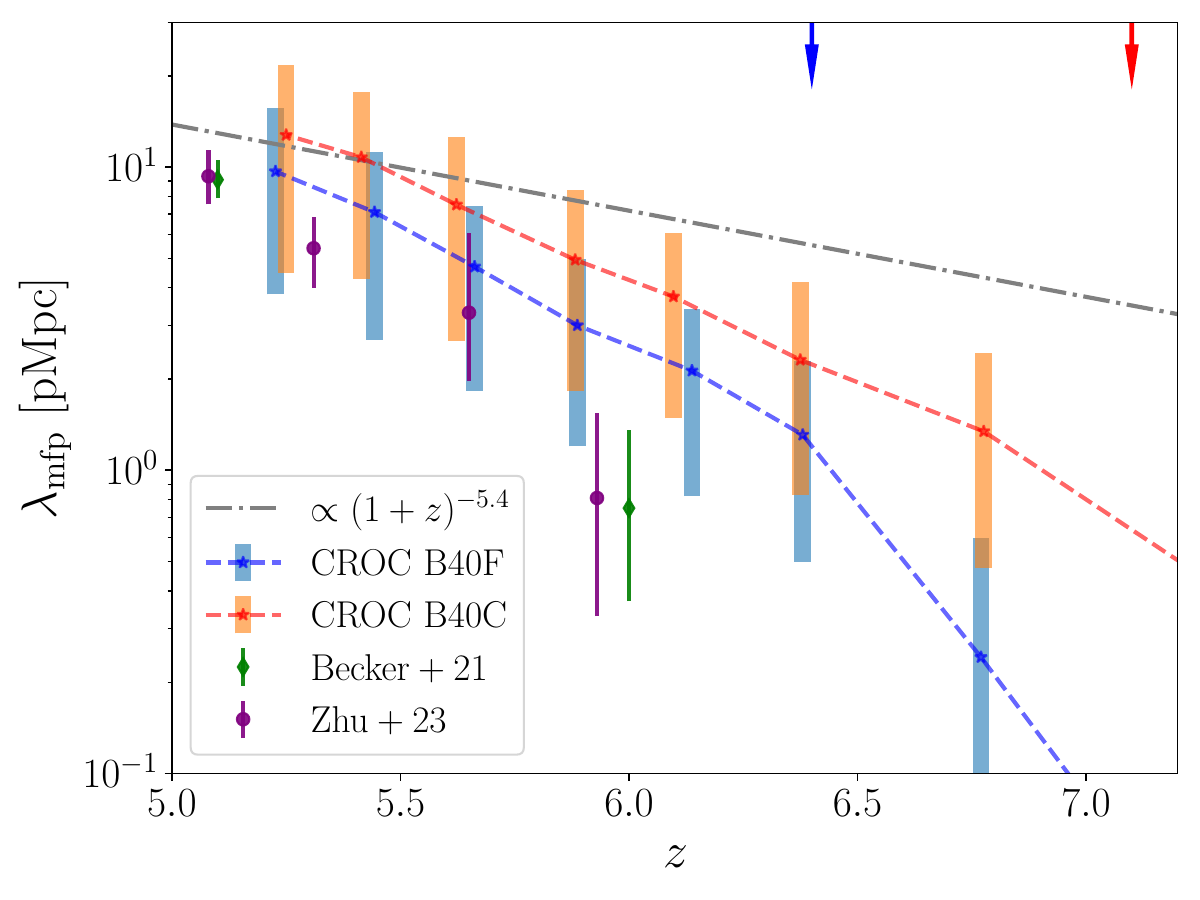}
    \caption{{Comparison of CROC mean free path evolution with previous work.} There is a break in the observed MFP evolution, similar to those in our CROC measurements that coincide with our measured `ankles' in the CROC reionization histories (indicated by the arrow ticks).  The location of the observed break hints at a later reionization history than what CROC captures.  But, direct interpretation of the comparison requires an accounting of quasar proximity zones.  }
    \label{fig:observational_overplot}
\end{figure}

In Figure~\ref{fig:observational_overplot}, we show observational measurements of the MFP overlaid on the CROC results. The CROC MFP is the same as Figure \ref{fig:mfp_evo}, except we only show the $68\%$ scatter about the median here for simplicity.
The recent measurements of MFP from \citet{Becker2021} and \citet{Zhu2023} are shown in green and purple, respectively. Each of their measurements is based on $\approx 10 \sim 40$ high-resolution quasars at that redshift, and the error bar represents their bootstrap error.
The grey dashed line shows the power-law relation $\lambda_{\mathrm{mfp}}\propto (1+z)^{\eta}$, where $\eta=-5.4$. This relation is the fitting result from \citep{Worseck2014} where they find that $\lambda_{\mathrm{mfp}}$ monotonically decreases smoothly within the range of redshift $z = 2.3$ and $z = 5.5$. At higher redshift, neither the observational measurements of \citet{Becker2021} and \citet{Zhu2023} nor the evolution we measure in the CROC simulation follow the power-law relation from the low redshift fit.

Instead, both the CROC B40F box (with the latest reionization history in the simulation suite)  and the observation results show a significant drop in MFP, creating a break in the evolution. Note, CROC B40F exhibits a break in a slightly higher redshift ($z\approx6.4$). Recall, this break in the CROC MFP evolution coincides with the when the last IGM neutral patches are ionized, or the `ankle' of the box neutral fraction evolution.  Considering the similarity between the CROC B40F and observed evolution of the MFP, this comparison to observational measurements may suggest that the ionization of the last patches of neutral IGM in our universe disappeared at $z\approx 5.6\sim 6$.  However, we caveat this suggestion with the fact that observational measurements are model dependent and quasar proximity effects have high uncertainties.  These uncertainties include the number of LLSs and DLAs surrounding the quasar, the quasar luminosity, and quasar lifetime. To correctly interpret the observational data, we need to thoroughly account for all these effects (Chen+2024 in prep, see also \citet{Satyavolu2023,roth2023}).

\section{Conclusion and Discussions}\label{sec:conclusions}
In this study, we analyze CROC simulations to gain insights into the co-evolution of the ionizing background and mean free path (MFP) during the late stage of reionization. The high resolution radiative transfer hydrodynamic simulations allow us to study these quantities and the effect of environment and dense clumps in the MFP measurement and evolution. We examine these relationships in the medium-sized CROC simulation runs with the latest (B40F) and earliest (B40C) reionization histories.  Each reionization history displays a knee,
where the volume-weighted neutral fraction begins to rapidly decline, and an `ankle' signifying the disappearance of neutral patches and a volume-weighted neutral fraction of $\xHIv\sim1\times 10^{-3}$.  The `ankles' of the B40F and B40C reionization history respectively occur at $z\sim6.4$ and $z\sim7.1$ (see Figure~\ref{fig:reihist}). We summarize our findings as follows.

\begin{itemize}
    \item The self-consistent accounting of reionization in CROC leads to a background radiation that exhibits a distinct skewness (Figure~\ref{fig:gamma_scatter_box_F} and \ref{fig:gamma_scatter_box_C}) that deviates from log-normal after the `ankle' in the box reionization history.
    \item We measure MFP with two common definitions, one without frequency dependency and with  frequency dependency (Figures~\ref{fig:lls_mfp_interaction} and \ref{fig:compare_two_def}), and find them to be almost identical. 
    \item Both the late and early reionization boxes, respectively B40F and B40C, display a break in their MFP evolution that coincides with the `ankle' of their respective reionization histories at $z=6.4$ and $z=7.1$ (see Figure~\ref{fig:mfp_evo}).  The rate of increase of the MFP slows with the disappearance of neutral patches and the sharp change in volume-weighted neutral fraction at $\xHIv\approx1\times 10^{-3}$.
    \item We examine the effect of LLSs and the environments in our MFP measurements. Sightlines that start from massive halos, `quasar-like' sightlines, have a relatively truncated MFP due to the concentration of LLSs and DLAs within 1 pMpc of the halo (see Figure~\ref{fig:column_density_vs_distance_evo}). We find that removing LLSs from `quasar-like' sightlines leads to a $\approx 20 \%$ increase in MFP at $z\approx5$.  We also find that the MFP of sightlines that start from random locations have $\approx 10 \%$ longer MFP compared with those that start from massive halos at $z\approx5$ (see Figure~\ref{fig:mfp_different_environments}). 
\end{itemize}
Finally, we compare the evolution of the MFP in CROC boxes to measurements from observations.  We note that there is a similar break in the observed MFP evolution, but this occurs at a later redshift suggesting that observations are consistent with an even later reionization history than those captured in the CROC boxes.  However, a caveat of our MFP measurements in CROC is that our measurements do not account for quasar proximity zones.  In a follow-up work, we will fully explore the dependence of the evolution of the ionizing background radiation and the MFP on assumptions surrounding quasar proximity zones.  We emphasize that a direct interpretation of the observed evolution of the MFP requires an accounting of quasar proximity zones.

\section*{Acknowledgements}

The authors thank Nickolay Y. Gnedin for providing access to the CROC simulation project and for helpful comments on the manuscript.  JF and CA thank Hanjue Zhu for useful discussions during the early stages of the project.  JF and HC thank Yongda Zhu for helpful input during the early stages of the project.  
JF acknowledges early support from the Summer Undergraduate Research Fellowship from the Physics Department at the University of Michigan.
HC thanks the support by the Natural Sciences and Engineering
Research Council of Canada (NSERC), funding reference \#DIS-
2022-568580.
CA acknowledges support from the Leinweber Center for Theoretical Physics.  

\bibliography{main}{}
\bibliographystyle{aasjournal}

\appendix
\section{Note on the Spatial fluctuation of $\gbg$}

\begin{figure}
    \centering
    \includegraphics[width=0.45\textwidth]{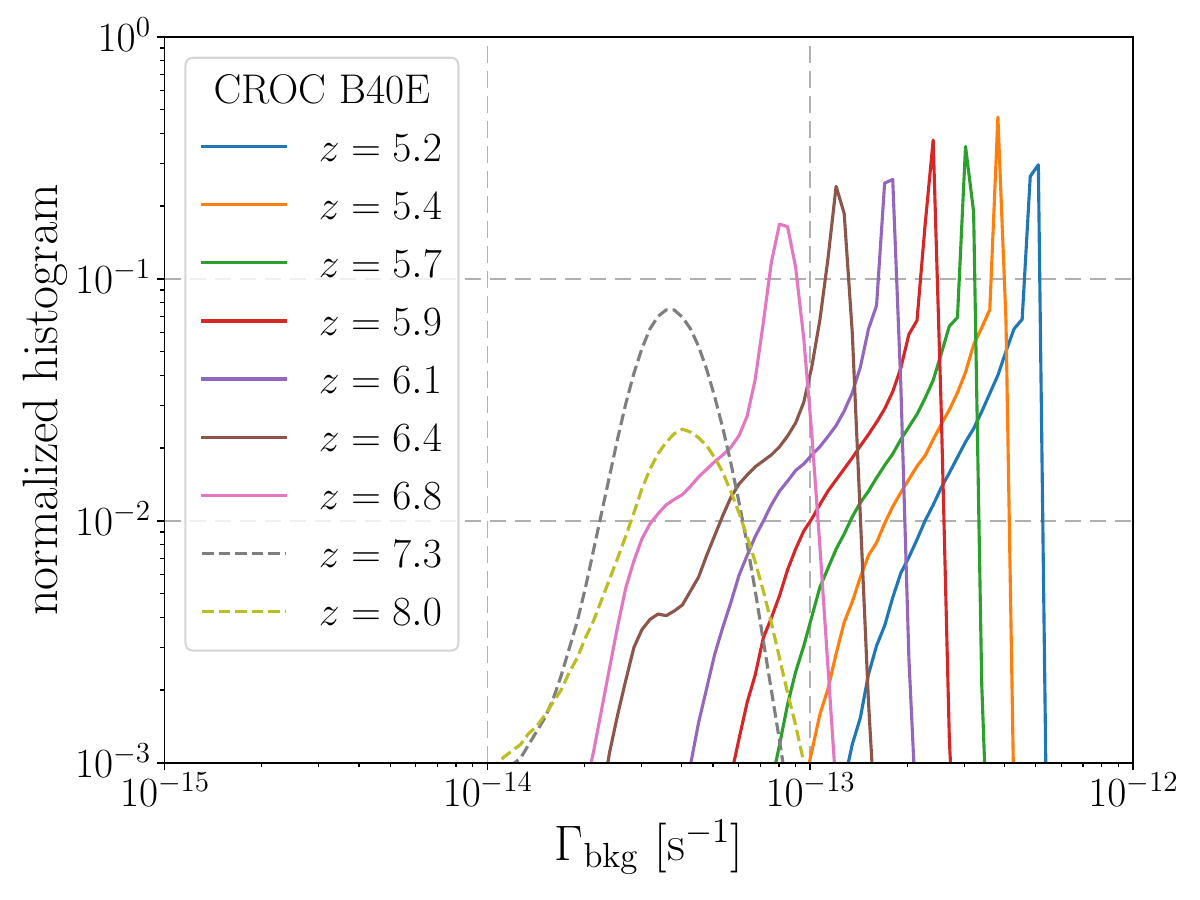}
    \caption{The $\gbg$ evolution of B40E, which has a reionization history in between the two boxes B40F and B40C we present in the main text.}
    \label{fig:gbg_B40E}
\end{figure}

The distinct `bump' feature in the $\gbg$ distribution in B40F is intriguing. 
It is likely that they correspond to the regions that are far away from ionizing sources and ionized the latest. This feature does not show up in the B40C box. The most plausible explanation is that in the overdense B40C box, ionizing sources are more numerous and there are no significant large regions that are far away from sources. To test this, we look into another CROC box B40E, which has a reionization redshift in between B40F and B40C. The $\gbg$ distribution of B40E (Figure \ref{fig:gbg_B40E}) behaves as we expected. The shape of B40E $\gbg$ distribution is in between those of B40F and B40C -- it has a wide width which resembles the `bump' of B40F, albeit with a steeper rise.

In addition, we visualize the snapshots in B40F and B40C to understand the physical nature of the regions connected to the `bump' features in the $\gbg$ histogram. For this purpose, we use uniform-grid data which is much more straight forward to visualize than the raw AMR data. Due to the fact that He neutral fraction is not saved, when calculating $\gbg$ using Eq. \ref{eq: bkg_gamma}, we further assume $n_e=1.1 n_{\rm HII}$. This is a fair assumption given the number density of He is around a tenth of H. 

\begin{figure*}
    \centering
    \includegraphics[width=0.32\textwidth]{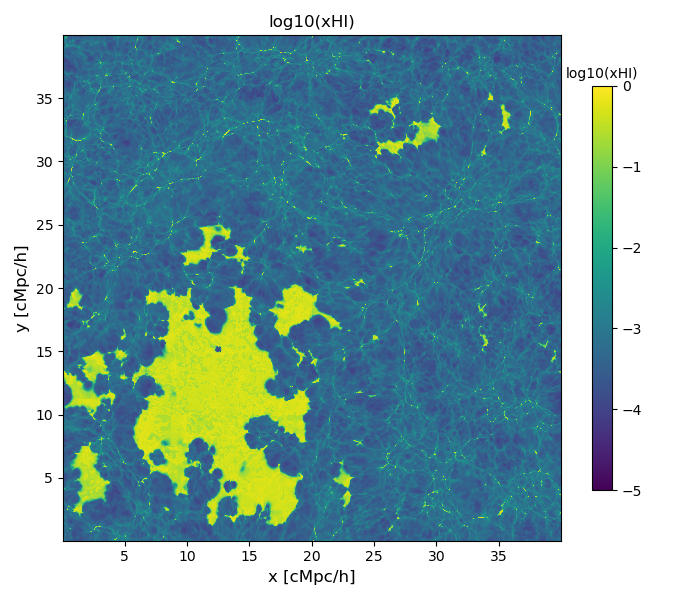}
    \includegraphics[width=0.32\textwidth]{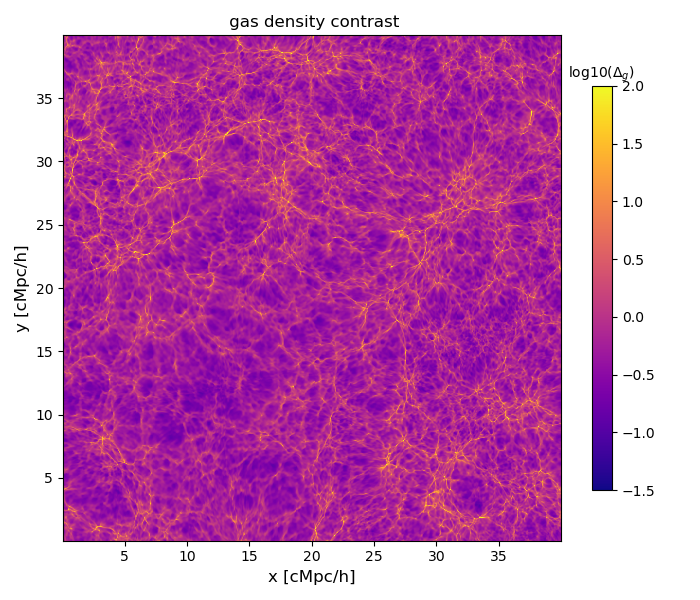}
    \includegraphics[width=0.32\textwidth]{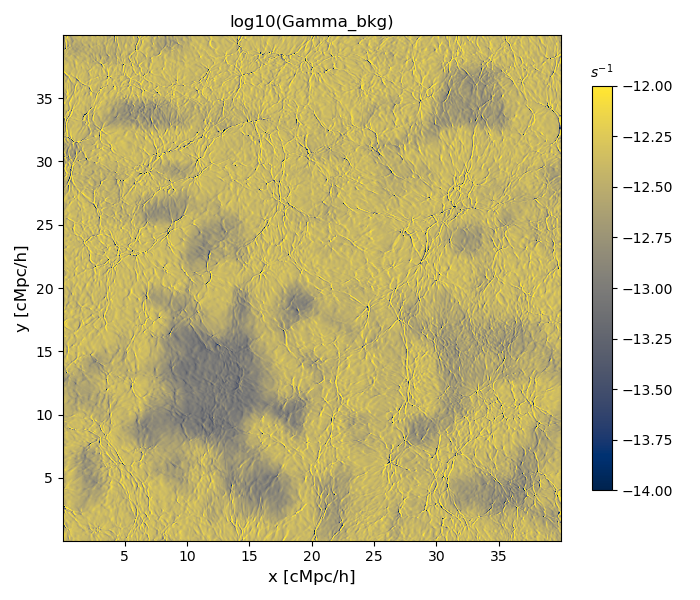}
        \includegraphics[width=0.32\textwidth]{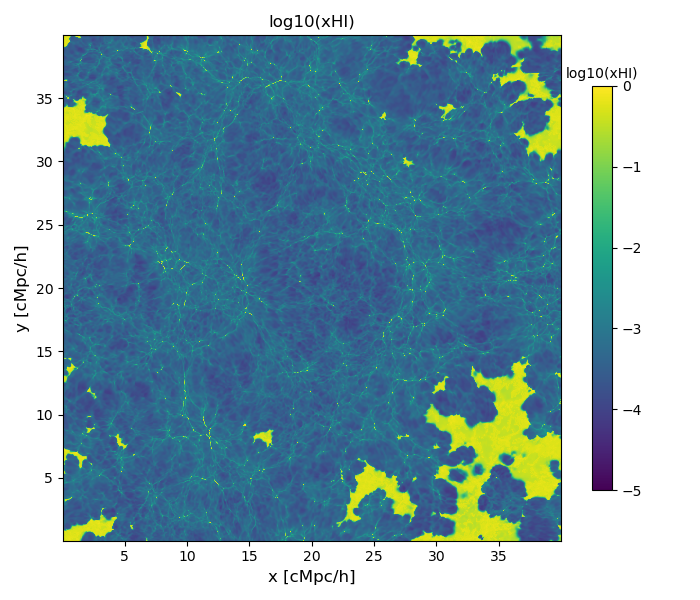}
        \includegraphics[width=0.32\textwidth]{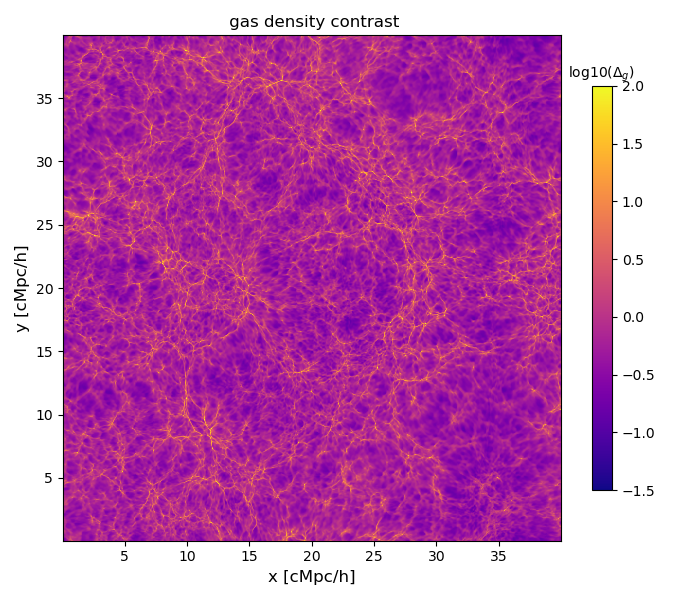}
        \includegraphics[width=0.32\textwidth]{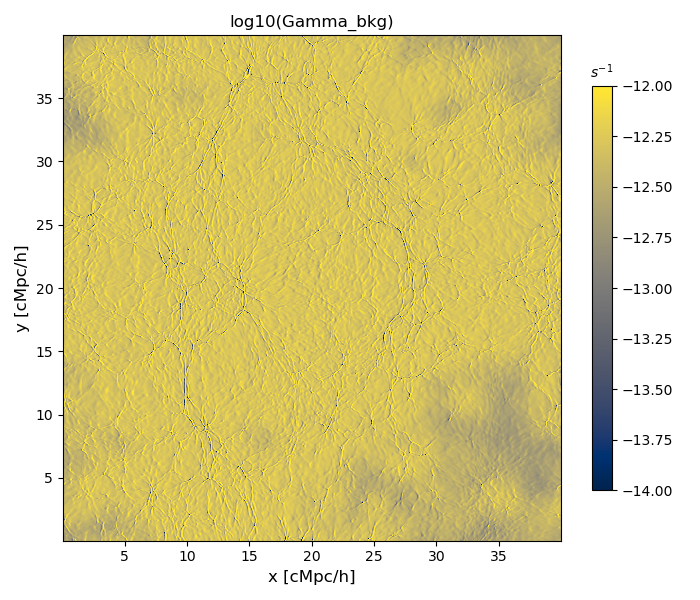}
    \caption{Maps of neutral fraction (left), density contrast (middle) and $\gbg$ (right) averaged over a thin slice of $39$ ckpc/$h$ in B40F (top row) and B40C (bottom row). The left column and middle columns are slices from epochs prior to when the last neutral patches disappear ($z=6.6$ for B40F and $z=7.2$ in B40C). The right column shows the $\gbg$ map well after the reionization of all IGM at $z=5.2$ for B40F and $z=5.4$ for B40C. The regions with lowest $\gbg$ corresponds to the largest and most underdense voids which contain the last patches of the neutral IGM.}  
    \label{fig:visualize}
\end{figure*}

In the first row of Figure \ref{fig:visualize}, we show a slice through the B40F box at $z=5.2$ (right panel) where there is a large patch where $\gbg$ has the typical values corresponding to the `bump' in the $\gbg$ histogram. We then investigate the same slice at a higher redshift of $z=6.6$, prior to the `ankle' in the box neutral fraction history, and find that the large patch corresponds to the last reionized IGM patch and the largest and most underdense void. On the contrary, B40C does not have such a large void where ionizing sources are scarce. The largest neutral patch that is ionized the latest is shown in the bottom row. The neutral patch in B40C (bottom left) is not as round as that in B40F (top left). Rather, the morphology of the neutral patch in B40C is concave with many ionized bubbles breaking into the patch. Because of the numerous ionizing sources surrounding the void in B40C, the relative deficit and contrast of the $\gbg$ of this region (dark splotch in lower right panel) is not as significant as in box B40F (larger darker splotch in upper right panel).

\label{lastpage}
\end{CJK*}
\end{document}